\documentclass{aastex}

\setcitestyle{citesep={,}}
\def\ee#1{\ifmmode {} \times 10^{#1} \else ${} \times 10^{#1}$\fi}

\def\aboutless{$\simless$\kern.03em}

\def\aboutmore{$\simgreat$\kern.03em}

\providecommand{\e}[1]{\ensuremath{\times 10^{#1}}}

\begin{document}

\title{\bf {Carbon monoxide in the distantly active Centaur (60558) 174P/Echeclus at 6 au}}

\shorttitle{CO in Centaur Echeclus} 

\bigskip

\bigskip

\author{K. Wierzchos$^1$, M. Womack$^1$, G. Sarid$^2$}

\affil{$^1$Department of Physics, University of South Florida, Tampa, FL 33620, USA}

\affil{$^2$Florida Space Institute, University of Central Florida, Orlando, FL 32826, USA}




\section{Abstract}

(60558) 174P/Echeclus is an unusual object that belongs to a class of minor planets called Centaurs, which may be intermediate between Kuiper Belt Objects and Jupiter Family comets. It is sporadically active throughout its orbit at distances too far for water ice to sublimate, the source of activity for most comets. Thus, its coma must be triggered by another mechanism. In 2005, Echeclus had a strong outburst with peculiar behavior that raised questions about the nucleus' homogeneity. In order to test nucleus models, we performed the most sensitive search to date for the highly volatile CO molecule via its J=2-1 emission toward Echeclus during 2016 May-June (at 6.1 astronomical units from the Sun) using the Arizona Radio Observatory 10-m Submillimeter Telescope. We obtained a 3.6-$\sigma$ detection with a slightly blue-shifted ($\delta$v = -0.55 $\pm$ 0.10 km s$^{-1}$) and narrow ($\Delta$v$_{FWHM}$ = 0.53 $\pm$ 0.23 km s$^{-1}$) line. The data are consistent with emission from a cold gas from the sunward side of the nucleus, as seen in two other comets at 6 au. We derive a production rate of Q(CO) = (7.7 $\pm$ 3.3)\e{26} mol s$^{-1}$, which is capable of driving the estimated dust production rates. Echeclus' CO outgassing rate is $\sim$ 40 times lower than what is typically seen for another Centaur at this distance, 29P/Schwassmann-Wachmann 1. We also used the IRAM 30-m telescope to search for the CO J=2-1 line, and derive an upper limit that is above the SMT detection. Compared to the relatively unprocessed comet C/1995 O1 (Hale-Bopp), Echeclus produces significantly less CO, as do Chiron and four other Centaurs.

\vfil\eject

\section{Introduction}

Minor body (60558) 174P/Echeclus is enigmatic: a rare dual comet-asteroid that is sporadically active throughout its orbit. It has peculiar behavior that may provide important tests to Solar System models. When originally discovered by Spacewatch in March 2000, it was near aphelion at $r$ = 15.4 au from the Sun, and had an asteroidal appearance and orbit. Thus, it received the designation 2000 EC$_{98}$ (60558). Five years later, it underwent a major outburst and produced a sizable dust coma, and so it also received the short-period comet designation 174P/Echeclus \citep{iau06}. It is  mostly dormant, with four known instances of activity: 1) a dust coma is evident in 2000 January ($r$ = 15.4 au) pre-discovery Spacewatch images, 2) a massive outburst during 2005 December ($r$ = 13.1 au)\citep{cho06}, 3) a moderate outburst in 2011 May ($r$ = 8.5 au)\citep{jae11}, and 4) another moderate outburst in 2016 August ($r$ = 6.2 au) \citep{mil16}. 

Echeclus belongs to a population of Solar System minor bodies known as Centaurs, whose unstable orbits range between Jupiter and Neptune (5 -- 30 au), have small inclinations, and may be in transition between the outer and inner Solar System \citep{wei93,lev97,hor04,sar09}. Echeclus is one of the few Centaurs to have a relatively high eccentricity ($e$ = 0.44), and it is sometimes called a ``Jupiter-coupled'' object \citep{gladman08}. Its last perihelion in 2015 April brought it relatively near the Sun for a Centaur: $q$ = 5.8 au. Most of the known Centaurs are asteroids, but 10\% to 15\% exhibit some kind of cometary activity \citep{bau08}. The evolution of KBOs into Centaurs and then Centaurs into JFCs can be understood as a diffusion from the Kuiper Belt to the inner Solar System after being sequentially exposed to the gravitational influence of the giant planets \citep{lev97,tis03}. Thus, their measured properties are valuable for constraining Solar System formation models \citep{bar10,van14}. 

Echeclus exhibited unusual behavior during its second outburst, which increased in total visual magnitude from 21 to 14 and produced a large dust coma. Interestingly, optical CCD images showed that most of the coma was separated from the central nuclear condensation, and appeared more diffuse than most comae \citep{rous08,bau08}. The detached coma's distance from the nucleus changed over time (from 1 -- 9$\arcsec$) and even appeared to backtrack at one point, reminiscent of a satellite in a hyperbolic orbit \citep{cho06, wei06}. The overall angular extent of the detached coma was $\sim$ 2$\arcmin$ (projected to be 1,040,000 km across at the comet's distance), surrounding a significantly brighter area 12$\arcsec$ in diameter \citep{teg06}. Further analysis concluded that the detached coma may have originated from a fragment a few kilometers or less in diameter, possibly ejected by impact or explosive outgassing from the much larger primary nucleus \citep{rous08,fer09}. The fragment's true nature is still unknown and it has not been seen since late 2006, nor was it visible in pre-discovery images. It may have been an isolated event that subsequently disintegrated into smaller pieces or sub-fragments \citep{fer09}. 

After the second outburst dissipated, Echeclus showed no signs of coma for many years \citep{lor07,rous08,cho08}, until the third outburst in 2011, when it increased in total visual magnitude from $\sim$ 19 to 14 \citep{jae11,rou15}. The coma was not detached, and it reached a diameter of $\sim$ 35$\arcsec$, which corresponds to a projected distance of 180,000 km at Echeclus. 

The fourth outburst recently occurred on 2016 August 28, when Echeclus was $r$ = 6.2 au from the Sun, post-perihelion, with an r$\arcmin$ magnitude of = 15.2\footnote{r$\arcmin$ Sloan is a variant of the photometric R filter.}, and an asymmetric coma 14$\arcsec$ across \citep{mil16}. The coma was observed on multiple occasions, reaching a maximum size of $\sim$ 72$\arcsec$, or 280,000 km at Echeclus' distance, on 2016 September 12 as reported by the amateur community \footnote{http://lesia.obspm.fr/comets/lib/all-obs-table.php?Code=0174P\&y1=1908\&m1=01}. The coma persisted until the beginning of October, and to date, no secondary component is reported from this outburst. 

Optical and infrared measurements provided additional important information about Echeclus. Its diameter is large for a comet, but typical of Centaurs. The best value of its diameter is from \citet{duffTNOs14}, who derive D$_{Echeclus}$ = 64.6 $\pm$ 1.6 km using data from Spitzer and Herschel space observatories. Dust production rates were estimated to be $\sim$ 700 kg s$^{-1}$ during the second outburst, and only 10 to 20 kg s$^{-1}$ for the third outburst \citep{rou15}. Spectrophotometric measurements of the surface during a dormant stage provided colors of V-R = 0.47 $\pm$ 0.06, and a R-I of 0.50 $\pm$ 0.06, which are consistent with the dynamical model of Centaurs originating from the Kuiper Belt \citep{bau03}. The dust coma colors obtained two months after the massive second outburst were V-R = 0.50 $\pm$ 0.06 and R-I of 0.58 $\pm$ 0.05, which is slightly redder than the reflected nucleus colors. This was interpreted as evidence for the presence of large grains up to $\sim$ 3mm across, \citep{kol01,bau03,teg08}, and inconsistent with an impact origin for the second outburst \citep{bau08}. 

Developing comae at such large distances is unusual. Most comets are not active at heliocentric distances beyond $\sim$ 3 au, where the temperature is too low for the water ice to sublimate efficiently (e.g., \citep{kel13}). Nonetheless, many comets are observed to have outbursts, or other outgassing, and other volatiles have been suggested as drivers of distant comet activity. A phase change of amorphous-to-crystallized water ice near the surface and subsequent release of trapped highly volatile molecules is considered to be an important contributor to activity for some comets within 5 -- 8 au \citep{pri90,sar05}, possibly even as far as 16 au from the Sun \cite{guilbert12}.  If gaseous emission is identified in Echeclus, this can provide important clues to the nucleus' chemical composition. Some leading contenders are CO, CO$_2$, and CH$_4$, which are cosmogonically abundant with sufficient equilibrium vapor pressures to vigorously sublimate at even very large ($>$35 au) heliocentric distances. 

Other well-studied Centaurs include 29P/Schwassmann-Wachmann 1 (hereafter 29P), 2060 Chiron, 10199 Chariklo, 5154 Pholus, 165P/Linear, 39P/Oterma, 165P/Linear, 8405 Asbolus, and C/2001 M10 (NEAT), some of which also exhibit activity \citep{boc01,jew09,duffTNOs14}. Four Centaurs have additional unusual features: Chariklo and Chiron appear to have ring-like structures \citep{duffrings14,rup15,sic16}, and there are reports of Echeclus (as described earlier in this section) and 29P both having a secondary source of dust emission (and a second small CO outgassing region for 29P), possibly emitting from a chunk or fragment in the coma \citep{sta04,gun08,wom17}. In order to test physical and chemical models of Centaurs (and hence by extension models of KBOs and JFCs), it is important to determine whether gaseous activity is common amongst Centaurs.

Space-based observations show that CO$_2$ is the dominant volatile for most distant comets \citep{bau15,oot12,rea13}, and many models incorporate it as a likely driver of distant activity \citep{pri04,bar13}. Unfortunately, CO$_2$ is difficult to observe from the ground, and thus, we focused on carbon monoxide emission, which is readily observable in cold cometary gas through its rotational transitions \citep{cro83}. 

Observations of two other well known Centaurs, 29P and Chiron, have activity that can be explained by CO outgassing, possibly involving the crystallization of amorphous water ice \citep{sen94,wom99}. It is difficult, however, to draw general conclusions about Centaurs based on such limited data. Previous attempts to detect CO in Echeclus with the JCMT 15-m telescope in 2002 May ($r$ = 14.8 au) yielded an CO production rate upper limit of Q(CO) $<$  3.6\e{27} mol s$^{-1}$ near aphelion \citep{jew08,dra17}. This limit was interpreted as being incompatible with there being much solid CO ice on the surface, but does not constrain models much further. 

Echeclus passed through its perihelion in Apr 2015 at $r$=5.8 au and this provided an excellent opportunity to search for CO emission. In late spring 2016 it was $r \sim$ 6 au, which is $\sim$ 2.5 times closer to the Sun and Earth than the previous best measurement of CO 14 years earlier. Thus, the nucleus would be significantly warmer, and any gas coma would appear larger in the sky, both of which should be more favorable for detecting CO outgassing. Observing at this distance provides a good test of nucleus models, which predict that outgassing should occur if CO ice, or trapped CO gas, is relatively near the surface. Echeclus is also within the $\sim$ 5-8 au zone where insolation can readily trigger amorphous water-ice to undergo crystallization, which also might release trapped CO gases. CO emission has been observed in other comets at this distance, most notably C/1995 O1 (Hale-Bopp) and 29P (also classified as a Centaur) \citep{sen94,wom97,biv02,gun03}. Thus, we performed the most sensitive search to date for CO emission in a Centaur, Echeclus. Here we present a 3.6-$\sigma$ detection of CO emission, derive CO production rates, and discuss implications.


\section{Observations and Results}

We searched for CO emission in Echeclus via its J=2-1 rotational transition at 230.538 GHz with two instruments: the Arizona Radio Observatory (ARO) Submillimeter 10-meter telescope (SMT) and the Institut de Radio Astronomie Millimetrique (IRAM) 30-m facility. Access to both telescopes was granted via director's discretionary time. The SMT was used to collect spectra from 2016 May-June, which led to a CO detection. During the SMT time, there were no reports of optical observations by the amateur or professional community, thus its activity status is unknown. On 2016 August 28, the comet underwent its fourth observed outburst, and produced a dust coma. Its visible brightness increased from $\sim$ 18 mag to 15, which was maintained for several days. A dust coma appeared and remained until at least mid-October. After the outburst, we immediately requested target-of-opportunity time on the IRAM 30-m telescope to search for residual CO emission that might be associated with the outburst. Observing time was granted, but not until 24 days after outburst. At this time Echeclus still had a dust coma, but its visible magnitude had dropped to $\sim$ 17 (MPS 730323). Thus, we assume that the IRAM data were obtained when Echeclus was not outbursting, but possibly still outgassing at a lower level. For both telescopes, comet positions and velocities were obtained with the Jet Propulsion Laboratory (JPL)/HORIZONS ephemeris system. Observations and results for both observing runs are summarized in Table \ref{tab:obs} and explained below.

\subsection{ARO SMT 10-m}

Observations using the SMT were carried out toward Echeclus between 2016 May 28 and June 13 when the Centaur's heliocentric, $r$, and geocentric, $\Delta$, distances ranged from  $r$ = 6.12 -- 6.14 au and $\Delta$ = 6.72 -- 6.58 au. 
The dual polarization 1.3 mm receiver was used, with ALMA Band 6 sideband-separating mixers. The data were obtained in beam-switching mode with a reference position of +2$\arcmin$ in right ascension, and an integration time of 3 minutes on the source and 3 minutes on the sky reference position. The temperature scale for all SMT receiver systems,T$_A^*$, was determined by the chopper wheel method, where T$_R$=T$_A^*$/{$\eta_b$}, and $\eta_b$ = 0.74 at 230 GHz. The backends consisted of a 2048 channel 1 MHz filterbank used in parallel (2 x 1024) mode and a 250 kHz/channel filterbank also in parallel (2 x 250). The SMT has a half power beamwidth of $\theta$ = 32$\arcsec$ at 230 GHz, which corresponds to an average projected diameter of 160.000 kilometers at the comet's distance. 

The spectra provided equivalent velocity resolutions of 0.325 km s$^{-1}$ and 1.301 km s$^{-1}$ for the 250 kHz and 1 MHz resolutions, respectively. The 250 kHz filter was the highest spectral resolution available, but the line still may not be resolved, since CO emission in distant comets is known to be very narrow, often 0.2 - 0.5 km s$^{-1}$. The 1 MHz line was significantly broader than the expected CO linewidth and thus was not used in the analysis. 

Focus and pointing accuracy were checked periodically on nearby planets, and the pointing consistently had an uncertainty of $<$ 1$\arcsec$ RMS. During June 10-13, the observing run suffered bad weather conditions and instrumental problems, and thus data from this time period were excluded. Otherwise, the weather was mostly very good and there were very few technical issues. Only scans with a telescope system temperature, T$_{sys}$, under 500K were used, which excluded 15 scans, resulting in a total of 23 hours of scans accumulated on the source. Individual scans were checked to rule out possible bad channels at the position of the observed feature. 

During the 2016 May-June observations with the SMT, we detected the CO J=2-1 line in Echeclus, as shown in Figure \ref{fig:original}. The feature was present in both spectral polarizations, and was seen in the data throughout the observing period. This feature has a cumulative signal-to-noise ratio of 3.6-$\sigma$, with a line intensity of T$_A$ = 8.53 $\pm$ 2.71 mK and line area of T$_A$dv= 4.8 $\pm$ 2.0 mK km 
s$^{-1}$ if a gaussian fit is assumed.

\subsection{IRAM 30-m}

\begin{figure}
\epsscale{.9}
\plotone{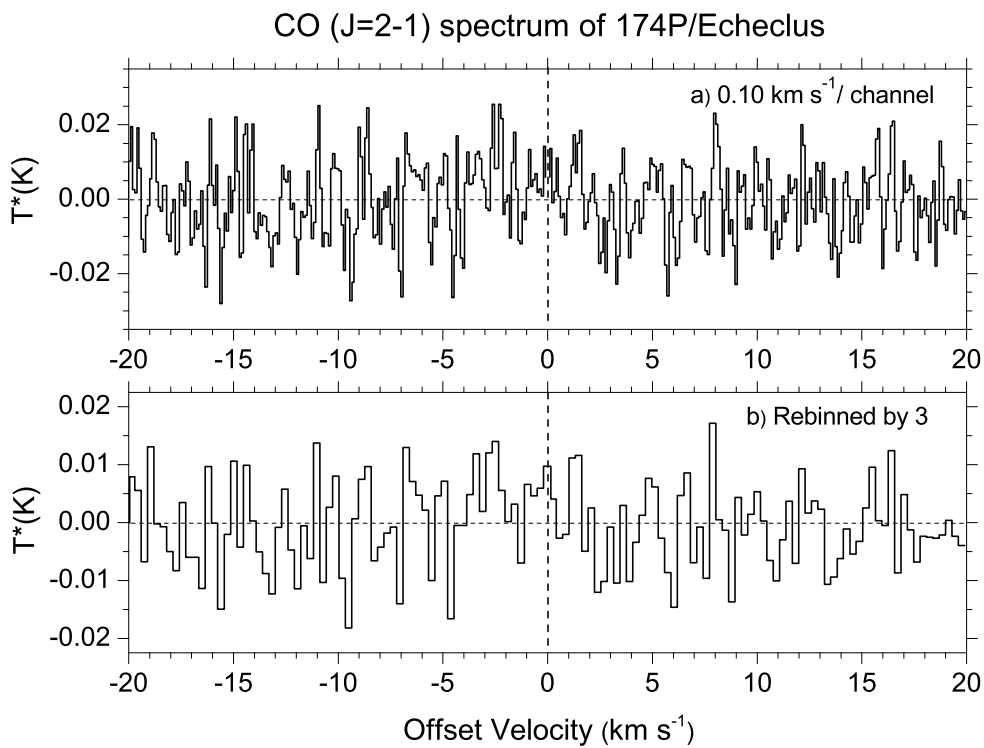}
\caption{
\label{fig:limit}
Spectrum of the CO J=2-1 rotational transition obtained with the IRAM 30m telescope on 2016 September 21. No line was detected on this date. The spectrum is presented with original spectral resolution corresponding to 0.10 km s$^{-1}$ and rebinned by 3 channels for an effective resolution of 0.31 km s$^{-1}$.} For comparison, Figures 1a and 2b have approximately the same spectral resolution.
\end{figure}

The IRAM observations were conducted during six hours on 2016 September 21, over a single transit, when Echelus was at $r$ = 6.31 au and $\Delta$ = 5.38 au. The measurements were made using the wobbling secondary mirror with a 2$\arcmin$ sky reference position, with the EMIR 230 GHz receiver, and high resolution VESPA autocorrelator at 80 KHz/channel resolution as backend.  The spectral resolution (0.104 km s$^{-1}$ when converted into Doppler velocity) is high enough to resolve most spectra from cold gas in distant comets. At this frequency, the IRAM half power beamwidth is $\theta$ = 10.7$\arcsec$, which corresponds to 41,000 km at the comet's geocentric distance on September 21.  The beam efficiency is $\eta_c$ =  0.59, and then the main beam temperature is T$_{mb}$=T$^*_A$/$\eta_c$. A total of 43 scans were obtained on source of 6 minutes each. Focus and positional accuracy were checked periodically on Uranus. The pointing consistently had an uncertainty of $<$ 1$\arcsec$ RMS. No line was detected with IRAM data (see Figure \ref{fig:limit}), and the noise RMS, uncorrected for beam efficiency, is 10.49 mK within $\pm$ 20 km s$^{-1}$  of the tuned frequency.


\section{Analysis and Discussion}

\subsection{Spectral line profile}

\begin{figure}

\epsscale{.9}

\plotone{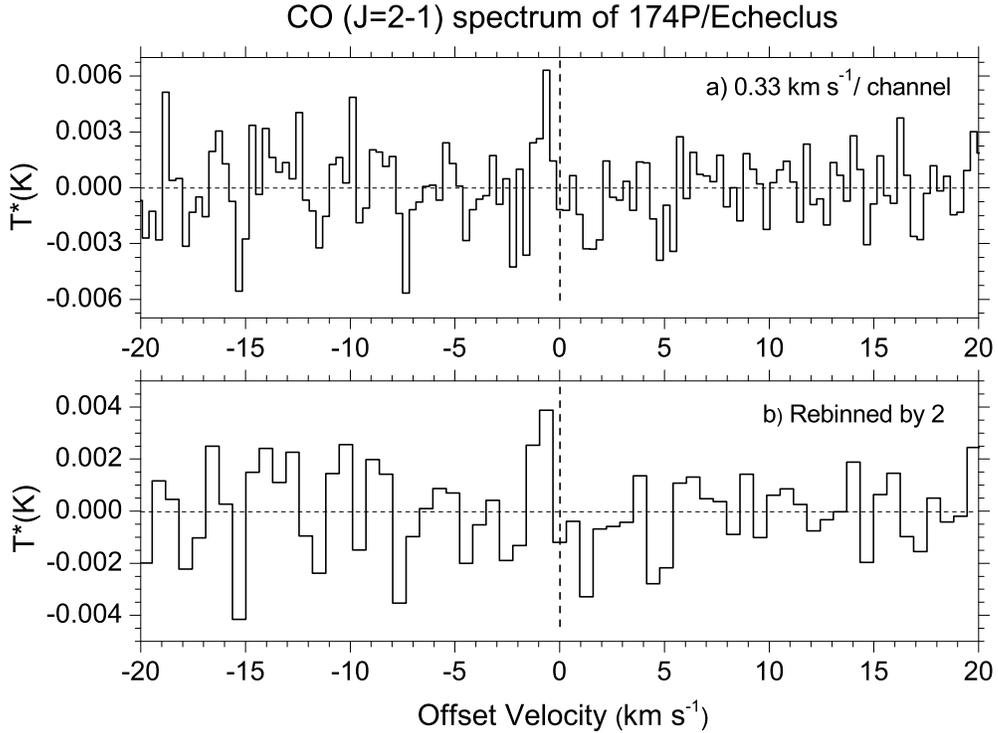}
\caption{
\label{fig:original} 
Spectrum of emission from the CO J=2-1 rotational transition obtained from the Arizona Radio Observatory Submillimeter 10-m telescope between 2016 May 28 and June 09. The spectrum is shown with original resolution equivalent to a) 0.33 km s$^{-1}$ and b) rebinned by 2 channels. The line has a cumulative signal-to-noise ratio of 3.6-$\sigma$, a FWHM line width of $\sim$ 0.53 km s$^{-1}$, and is shifted from the predicted ephemeris velocity (denoted by dotted vertical line) by -0.55 km s$^{-1}$. This line profile is consistent with emission of a cold gas from beneath the nuclear surface on the sunward side of the nucleus. 
}
\end{figure}

Spectral line profiles have characteristics, such as linewidths and velocity shifts, that provide important information about the gas emission release. A gaussian fit to the CO feature in the SMT spectrum yields a narrow line with $\Delta$V$_{FWHM}$ $\sim$ 0.53 $\pm$ 0.23 km s$^{-1}$ that is blue-shifted by a small amount from the expected comet ephemeris geocentric velocity, $\delta$v = -0.55 $\pm$ 0.10 km s$^{-1}$, see Figure \ref{fig:original}. The feature is not spectrally resolved, so the true linewidth could be narrower.
 
The blue-shifted, narrow line profile is very similar to what was observed at 6 au for CO emission lines in the comae of 29P and Hale-Bopp, and comparable to what was measured for Chiron at 8.5 au, which gives additional credence to the 3.6-$\sigma$ detection \citep{sen94,jew96,wom97,wom99,biv02,gun03,gun08}. It also suggests a common outgassing mechanism for CO for these objects. The line profile in Figure \ref{fig:original} is consistent with emission of a cold gas originating beneath the nucleus surface on the sunward side, and is also explained by hydrodynamical models which predict very low temperatures for cometary volatiles in the absence of photolytic heating \citep{cro95,sar09}. With respect to the observing geometry at this time, Echeclus' phase angle was very low, $\beta$ $\sim$ 8 degrees, which means that the sunward side was very similar to the Earthward side. A more detailed modeling of the line profile is beyond the scope of this paper.

\subsection{Excitation and production rates}

Using the ARO SMT spectrum, we derive a total column density of N(CO) = (4.9$\pm$ 2.0)\e{12} cm$^{-2}$ assuming both collisional and fluorescence excitation in a manner similar to \citet{cro83} and \citet{biv97}. We also assumed an optically thin gas and rotational and excitation temperatures of 10K and a gas expansion velocity of 0.2 km s$^{-1}$. These are consistent with the spectral line profile parameters described in the previous section and with values for CO emission in comets and Centaurs at large heliocentric distances \citep{biv02,gun08,dra17}. We calculate a production rate of Q(CO) = (7.7 $\pm$ 3.3)\e{26} mol s$^{-1}$, assuming isotropic outgassing of CO from the nucleus and a photodissociation decay model \citep{has57}. Applying a narrow cone ejection model, which may be justified by the blue-shifted spectrum consistent with sunward-side emission, reduces the production rate by ∼40\%. Our calculations assume that the CO emission fills the telescope beam.  A detailed study of CO  in comets showed that substantial amounts could be produced by extended sources within $\sim$ 2-3 au of the Sun \citep{pierce10}; however, because we are concerned with an object much farther out, we assume that all CO is from the nucleus. 

We did not detect a line with the IRAM 30-m data, and derive a 3-$\sigma$ upper limit to the line area by multiplying the RMS, T$^*_A$=10.49 mK, by three, correcting it for beam efficiency ($\eta_c$ = 0.59), and assuming the same linewidth  as the observed FWHM (0.53 km s$^{-1}$) from the SMT observations.  Using this upper limit for the linear area and the same modeling parameters as in the previous section, we calculate an upper limit of Q(CO) $<$ 12.2\e{26} mol s$^{-1}$ for 2016 September 21, when the comet was 6.3 au from the Sun. This limit is higher than the value we derived a couple of months earlier with the SMT (see Table \ref{tab:obs}).

\subsection{CO production in Centaurs}

Here we briefly summarize what is known about CO production rates from Centaurs. Carbon monoxide emission is now reported for three Centaurs: 29P, Chiron and Echeclus. Since its first cometary detection in 1994, CO is repeatedly seen in 29P, an unusual object that has a nearly circular orbit, ranging from $r$ = 5.7 to 6.3 au and is continuously outgassing \citep{sen94,gun08}. In contrast, CO emission was marginally detected in Chiron at 8.5 au \citep{wom99}, and now in Echeclus at 6.1 au, both at much lower production rates than 29P.\footnote{We note that when subsequent searches did not detect CO in Chiron over $r \sim$ 9 - 11 au, questions were raised about whether the Chiron CO detection was real \citep{boc01}; however, an independent re-analysis of the original data maintained that the CO line was formally detected \citep{jew08}.} A recent study set Q(CO) upper limits for 16 Centaurs (including Echeclus at aphelion) and analyzed them with a high-albedo energy balance model \citep{dra17}. They ruled out substantial amounts of CO ice being present on their surfaces, but did not constrain models much further. 

Our measured value for Echeclus of Q(CO) =  7.7 \e{26} mol s$^{-1}$ is the lowest measured Q(CO) in any Centaur, and is $\sim$ 5 times lower than the upper limit of  Q(CO) $<$ 3.6\e{27} mol s$^{-1}$ derived for Echeclus in the earlier study. This new result compels us to revisit the Centaur outgassing behaviors.
An interesting point can be made if we bring in a large (non-Centaur) comet into the discussion. Although Hale-Bopp is not a Centaur, we include it for reference as a relatively unprocessed comet (it is a long-period comet and thus spends most of its time at very large heliocentric distances), and because CO was observed throughout its orbit \citep{biv02}. Echeclus' CO production rate is $\sim$ 10-50 times lower than what was reported for both 29P and Hale-Bopp at $r \sim$ 6 au. The significantly lower CO output of Echeclus, compared to 29P and Hale-Bopp is noteworthy, since all three have similar sizes and the measurements are all at $\sim$ 6 au from the Sun. Echeclus is approximately the same diameter (D$_{Echeclus}$ $\sim$ 65 km) as Hale-Bopp (D$_{HB}$ $\sim$ 60 km,  \citet{fer00}) and 29P (D$_{29P}$ $\sim$ 60 km, \citet{sch15}). This fortuitous alignment of size and heliocentric distance allows us to eliminate some key parameters when comparing their CO production rates. Thus, Echeclus clearly produces CO at a much lower rate than both 29P and Hale-Bopp. We cannot tell whether this is due to a difference in composition from different formation environments, or whether Echeclus originally had more CO, and is now devolatilized. Since Echeclus and 29P are both considered to be Centaurs, this difference in CO output is especially striking and may be useful for constraining Centaur models. Similarities and differences of CO outgassing from 29P, Hale-Bopp and Chiron are discussed in more detail in \citet{wom17}.

\subsection{CO production rates adjusted for nucleus size}

\begin{figure}
\epsscale{.9}
\plotone{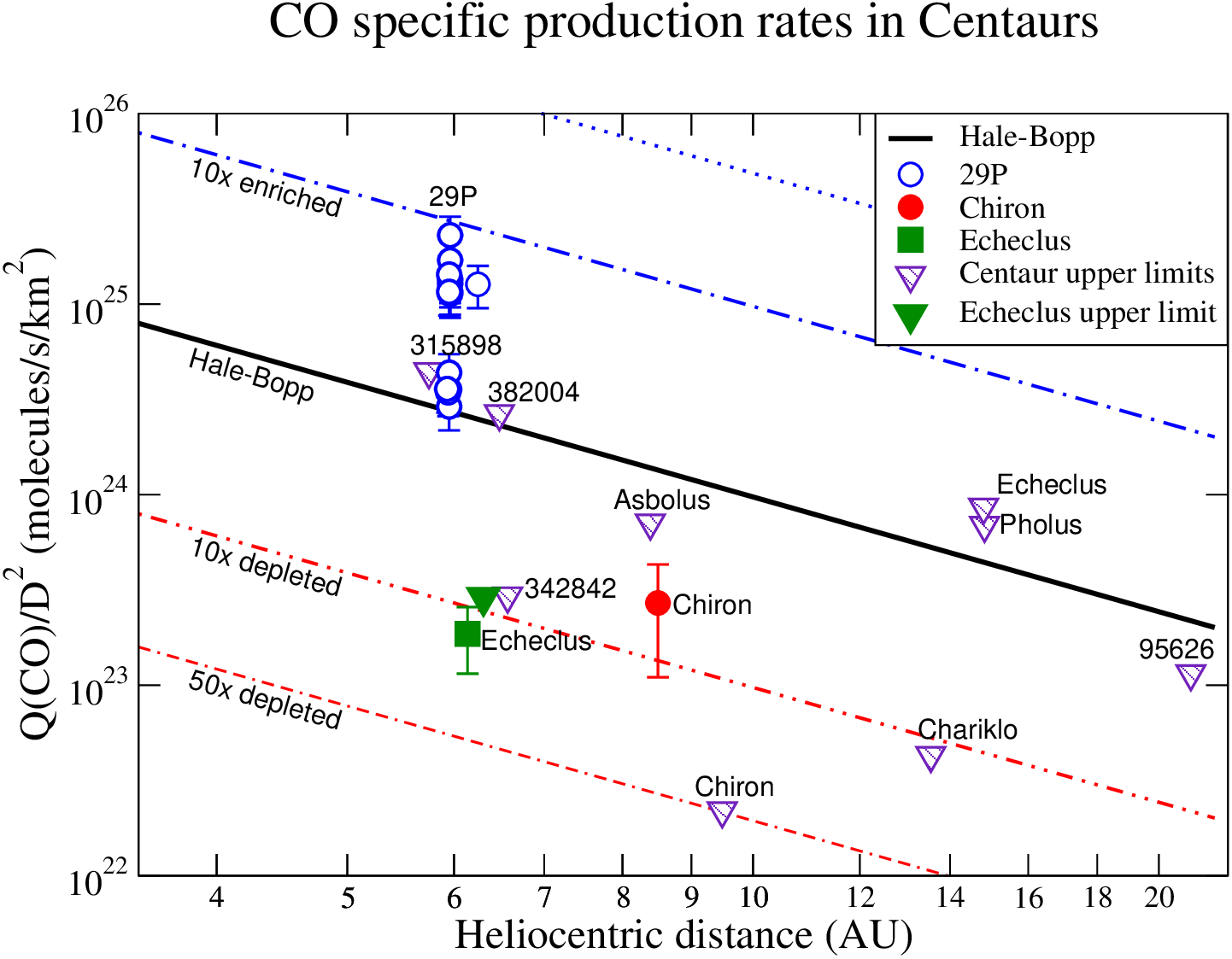}
\caption{
\label{fig:specific}
Specific production rates, Q(CO)/D$^2$, for distant comets and Centaurs. The solid line is 
a fit to data for Hale-Bopp, assuming Q(CO)=3.5\ee{29}$r^{-2}$, and diameter 
D$_{Hale-Bopp}$=60 km; values for Chiron and 29P are plotted assuming production rates from 
\citet{wom17} and D$_{Chiron}$=218 km and D$_{29P}$=60 km. This plot
shows that, when adjusted for surface area and heliocentric distance, Echeclus (D$_{Echeclus}$=65 km) and Chiron produce $\sim$ 10-50 times less CO than Hale-Bopp, a proxy for a relatively unprocessed comet. Upper limits to specific production rates for other Centaurs were calculated from Q(CO) upper limits and diameters provided in \citet{dra17}, and show that 29P is the only CO-rich Centaur found to date. All Centaurs with values below the Hale-Bopp line are considered to show a reduced CO activity level.}
\end{figure}

To explore Centaur outgassing further, we consider CO output in other Centaurs adjusted for nucleus size. If we assume that Centaurs and comets have similar chemical and physical compositions, then their CO production rates should scale commensurate with surface area and distance from the Sun. Thus, we calculated the specific production rate, Q(CO)/D$^2$, of the Echeclus detection and upper limit, and plotted the results in Figure \ref{fig:specific}, along with values of Q(CO)/D$^2$ for Hale-Bopp, following the method we introduced in \citep{wom17}. We also calculated upper limits to specific production rates for several other Centaurs using measurements from \citet{dra17}, which are also included in the figure.

Following this line of reasoning, many Centaurs appear CO-poor when compared to comet Hale-Bopp. For example, Echeclus and Chiron both appear to produce 10-50x less CO/area than Hale-Bopp and 29P. We think that the Echeclus and Chiron detections are credible; however, even if one instead conservatively considers these CO measurements as upper limits, then the conclusions still stand. In fact, if these were only upper limits, then Echeclus and Chiron would produce even less CO than inferred from the detections. When we include the upper limits derived from \citet{dra17}, even more Centaurs appear deficient, or depleted, in CO and may, in fact, be inactive. As Figure \ref{fig:specific} shows, other Centaurs, such as Chariklo and 342842, are each CO-depleted by $\sim$ 10x compared to Hale-Bopp at the same heliocentric distance. The upper limits for 8405 Asbolus and 95626 are consistent with objects that produce at least 5x less CO/area than Hale-Bopp. Note that all the values we use from \citet{dra17} are upper limits, so the true CO output from these objects may be far lower, or they may be totally inactive. It is difficult to interpret an upper limit for non-detection of gas emission,  even more so if the objects are not known to display activity, as measured in gas or dust emission, as is the case for Chariklo and Pholus. Outer solar system objects that may appear depleted in volatile species, due to lack of gas activity detection, can still host a variety of volatile ices in their sub-surface layers. This is due to the difference in sublimation curves between volatile species and the specific physical and chemical histories of each object, which can include origin location in the disk, orbital variations and heating episodes (insolation, impact, radioactive). Thus, lack of gas emission, or even detection of lower levels, may not indicate depletion in an object's volatile species reservoir, but rather a suppressed level of measurable activity (referred to as “quenched activity” in the literature, e.g., \citet{pri04}).

	This effective quenching of the object's activity may be due to formation of lag deposit and subsequent dusty crust, which inhibits the heat transfer and the gas permeability of the upper layers, reducing the sublimation rate of volatile species and the flow of such gases to the surface. Another possibility is that past activity, either under current conditions or at different heliocentric distances, caused desiccation of the object's outer layers with little to no accessible volatile species surviving (e.g., \citep{pri04,meech04}. When we look at gas activity measurements, depletion of volatile species, in terms of bulk composition ratios, may not be functionally distinct from those mechanisms of mantling or desiccation. However, we suggest that CO may be present at higher levels, at deeper sub-surface layers, but suppressed in detectable emissions.

	Small bodies in the outer solar system present a real challenge in determining compositional trends. Unlike the growing chemical database for comets \citep{cochran15}, the more distant and less active Centaurs lack sufficient information, either from surface reflectance or from gas emission, to determine mixing ratios between volatile species. Thus, we make the distinction that depletion in these cases refers to the reduced activity level detected for a given species (such as CO) relative to other objects or a certain reference. That reference in our case here is comet Hale-Bopp. This is analogous to the taxonomy framed for highly-active objects \citep{mum11}. We note that in the absence of complete knowledge of mixing ratios (at least relative to water), it is complicated to distinguish between different ratios of carbon species in the nucleus and production rates of molecules in the coma overall \citep{sar05}.           

Interestingly, with the notable exception of 29P, no Centaurs produce the same amount of CO per surface area as Hale-Bopp. Thus, there may be a significant difference in composition and/or bulk properties between most Centaurs and Hale-Bopp. We note that most of the Centaurs below the line in Figure \ref{fig:specific} have the largest nuclei. As discussed in \citet{wom17}, CO devolatilization may be expected in large Centaurs, partly due to increased radiogenic heating. It is important to keep in mind, however, that this is partly a selection effect, because only significant limits have been made to date only for Centaurs with diameters greater than 65 km. We point out that 29P, the most CO prolific Centaur, is on the small size for Centaurs with a diameter of $\sim$ 60 km. It will be very useful to obtain more constraining Q(CO) measurements on smaller Centaurs. It may also be helpful to extend this analysis to other comets; however, we were unable to find any other distant comet Q(CO) measurements beyond 4 au which had independently determined Q(CO) and diameters. 

Although Centaurs are thought to share a common origin in the Kuiper Belt, the well-known different outgassing behavior in Chiron and 29P has been attributed to the differences in composition and evolution history \citep{cor08,des01}. The apparent CO deficiency in some Centaurs, consistent with Figure \ref{fig:specific}, may be explained by the devolatilization of small KBOs, in which only large KBOs are able to retain volatiles since the rate of volatile loss is controlled by the gravity and surface temperature \citep{sch07,bro12}. 

We bring up another potentially important characteristic: activity rate. Chiron and Echeclus (both very low CO outgassers) outburst only rarely, often with years between outbursts; whereas, 29P (CO-rich) has multiple outbursts per year. Since Echeclus and Chiron are only occasionally active, one possible cause is that emitted gas (possibly CO) originates from sub-surface ice patches, or possibly as trapped volatiles that are freed by the crystallization of amorphous water ice. Echeclus and Chiron may have had more CO in the past and substantially devolatilized after moving into their Centaur orbits.  In contrast, we note that 29P undergoes an outburst every $\sim$ 50 days on average.  29P looks like an outlier Centaur, by producing so much CO. Perhaps it is a more recent entrant to its current orbit than frequently thought, or perhaps it has a different origin from most Centaurs and had more CO incorporated during its formation. 

Whether CO or another volatile is dominant in Echeclus or Chiron, we cannot tell, but we note that the measured CO production rates are high enough to explain measured dust production rates \citep{rous16}. Other highly volatile and abundant gases, such as CO$_2$, CH$_4$ and N$_2$ may contribute to distant activity, but they are difficult to observe and none have been detected in any Centaur to date. N$_2$ abundances 
can be estimated indirectly from optical spectra of N$_2^+$, and so far all data point to N$_2$ $<<$ CO in comets \citep{wom92,lutz93,coch00,coch02,iva16}. Sublimation temperatures for these other volatiles are low enough that they may be expected to start sublimating if any pockets (localized solid concentrations) are sufficiently close to the surface. Thus, the presence of these species in the nucleus cannot be ruled out.

\subsection{Outbursts, fragmentation, and transient material} 

We briefly turn our attention to four Centaurs known to have additional unusual features, as described in the introduction. It is interesting that fragments, rings and ring arcs have been reported very near, or even in orbit around these Centaurs: Chariklo, Chiron, 29P and Echeclus. The origins of these features are still unknown. It is further interesting to postulate a correlation between the CO-activity level and the mass and distribution of these secondary components. One possible explanation is that perhaps these Centaurs' nuclei are more inhomogenous in their highly volatile ices, namely CO, and thus they might be more prone to eject chunks of various sizes, as a strong outburst can release large fragments, whereas weaker outgassing can expel smaller particles \citep{panwu16}. Then, due to their mass and gravitational pull, the lofted material can remain in semi-bound or bound orbits around the primary. Subsequent evolution into the features we observe today (fragments, rings, arcs) should be studied more carefully, as well as its correlation with actual observed activity. We leave this for future work. 

If CO is present in Echeclus and other Centaurs, and is released when amorphous water ice becomes crystallized, then amorphous water-ice should be present in their KBOs progenitors and in their possible descendents, the JFCs. However, thus far, the ice found in several KBOs, like Charon or Quaoar, is crystalline and not amorphous \citep{jew08}. These studies cannot rule out the existence of some patches of amorphous water ice on, or just below, the surface. Such patches might reach the critical temperature once the object comes close enough to the Sun. The behavior of Centaurs displaying volatile outbursts when pulled into the inner Solar System from the Kuiper Belt region was found to be consistent with a devolatilized KBO progenitor with a compositionally layered structure \citep{des01}. 
Alternatively, the low CO outgassing observed in Echeclus may be due the presence of a crust on their surface, making it difficult for the volatile molecules to flow through the dust external layers  \citep{cor08}.  This last scenario is considered unlikely for Echeclus, due to the fact that after the second outburst and fragmentation, virtually all the activity was from the detached coma, with very little, if any activity coming from the newly exposed inner regions \citep{rous08}. 

Further observations and modeling of other Centaurs, especially active ones, could provide important clues to the bulk properties of their nuclei and place powerful constraints to Solar System models. In particular, further observations of Echeclus are needed as it moves away from perihelion over the next few years.

\section{Conclusions}

We report a 3.6-$\sigma$ detection of the CO J=2-1 emission line in Centaur 174P/Echeclus during 2016 May-June with the SMT 10-m telescope, which yields to a Q(CO) production rate of (7.7 $\pm$ 3.3)\e{26} mol s$^{-1}$, when the object was at $r$ = 6.1 au and $\Delta$ = 6.6 au. This is the lowest measured Q(CO) in any Centaur, and is $\sim$ 5 times lower than the upper limit of  Q(CO) $<$ 3.6\e{27} mol s$^{-1}$ derived from data obtained near aphelion \citep{dra17}. No line was detected 24 days after the 2016 August 28 outburst using the IRAM 30-m, but a 3-sigma upper limit of Q(CO) $\textless$ 12.2\e{26} mol s$^{-1}$ was inferred at $r$ = 6.3 au, which is consistent with our SMT measurement. 

Our data are consistent with Echeclus being a CO-deficient body, when comet Hale-Bopp's data at $\sim$ 6 au is used as a proxy for a relatively unprocessed nucleus. When scaled by surface area and compared with CO measurements from other Centaurs, we see that specific production rates, Q(CO)/D$^2$, from Echeclus and Chiron are $\sim$ 10-50 times below that of Hale-Bopp, and 29P, which are both known to be CO-rich. The lower CO output of Echeclus and Chiron may mean that they incorporated less CO into their nuclei than many other comets, or they may have lost a significant amount due to devolatilization while in their relatively close-to-the-Sun Centaur orbits. It is puzzling why 29P, another Centaur, is evidently CO-rich.  Although no other CO detections exist for other Centaurs, stringent upper limits to their specific production rates show that several other Centaurs are also notably absent of CO outgassing, including Chariklo, 8405 Asbolus, 34842 and 95626. 

It is noteworthy that fragments, rings and ring arcs have been reported very near, or even in orbit around four Centaurs: Chariklo, 29P, Chiron, and Echeclus. The origins of these features are still unknown, but three of them (29P, Chiron and Echeclus) are active. Further measurements of CO production rates and documentation of any fragmentation in Echeclus and other Centaurs are needed to constrain models of the bulk structure and strength of Centaurs, their KBO predecessors and JFC successors.

\section{Acknowledgements}

We thank the ARO 10m and the IRAM 30m telescope crew and appreciate the help of Albrecht Sievers. The SMT is operated by the ARO, the Steward Observatory, and the University of Arizona, with support through the NSF University Radio Observatories program (AST-1140030). MW also acknowledges support from NSF grant AST-1615917. This research has made use of data and/or services provided by the International Astronomical Union's Minor Planet Center. This work is based in part on observations carried out under project number D07-16 with the IRAM 30m Telescope. IRAM is supported by INSU/CNRS (France), MPG (Germany) and IGN (Spain).

\vfil\eject
\bibliography{BigReferences}
\bibliographystyle{aasjournal}

\vfil\eject

\begin{table}
\epsscale{.9}
\caption{Observations of CO J=2-1 line in 60558 Echeclus}

\begin{tabular}{llrrll }\hline

UT Date (2016) & Telescope & $r$ (AU) & $\Delta$ (AU) & Tadv (mK km s$^{-1}$)& Q(CO)(\ee{26} mol s$^{-1}$)  \\\hline

May 28 -- Jun 09 & SMT 10m & 6.1 & 6.6 & 4.8 $\pm$ 2.0 & 7.7 $\pm$ 3.3  \\

Sep 21 & IRAM 30m & 6.3 & 5.3 & $<$ 28 & $<$ 12.2  \\
\hline
\end{tabular} 
\label{tab:obs}
\end{table}

\vfil\eject

\vfil\eject

\vfil\eject

\end{document}